# Entanglement distillation for three-particle W class states

Zhuo-Liang Cao[*], Ming Yang

Department of Physics, Anhui University, Hefei, 230039, PRChina

## Abstract

In this paper, we propose two general entanglement distillation protocols, which can concentrate the non-maximally entangled pure W class state. The general protocols are mainly based on the unitary transformation on the auxiliary particle and one of the three entangled particles, and in the second protocol, the entanglement distillation includes two meanings, namely, extracting the concentrated tripartite entangled W state and obtaining the maximally entangled bipartite state from the *garbage* state, which gives no contribution to the distillation of non-maximally entangled pure W class state. We can make use of the *garbage* in the distillation process, and make the entanglement waste in quantum communication as small as possible. A feasible physical scheme is suggested based on the cavity QED.

PACS: 03.67.Hk, 03.65.Bz

## Key words

W state, entangled atomic states, entanglement distillation protocol, Cavity QED

---

[*] E-mail address: caoju@mars.ahu.edu.cn

# I    Introduction

The most important and intriguing feature of entanglement is the non-locality property, which has been clearly illuminated by the Einstein, Podolsky and Rosen paradox [1]. When one particle of a two-particle entangled system was sent to a distant location (A), the other to another location (B) in an opposite direction, there are some subtle connections between the two particles because of entanglement. Before measured, the state of every particle is determined by the other one's, that is to say, the state of one particle is suspended between two different states (here, we have supposed that the particles are two-state systems). If you have measured the particle in location A, the particle in location B will inevitably collapse into one certain state, whether Bob(the receiver in location B) measures it or not. That is the so-called non-local connection. This novel property makes the entangled state a critical source for the quantum communication [2]. So quantum communication makes it possible to transmit an unknown quantum state without the transmission of the carrier (for instance, atoms and photons *et al*) of the unknown state itself. In recent years, there has been a rapid improvement in quantum communication [3-9]. If the quantum channel is a maximally entangled state, the fidelity of the transmission can reach 1.0[2], but the entangled states distributed among distant locations are usually non-maximally entangled, which is resulting from different noises or non-optimal preparation scheme. So some other probabilistic teleportation schemes have been proposed [10-13]. In another way, if we extract the maximally entangled states from the non-maximally entangled states, we can, consequently, realize the quantum teleportation with fidelity 1.0. Thus, the discussion about the entanglement purification is of practical significance.

A main obstacle of the long distance quantum communication is the decoherence of entanglement shared by different users. Because we can't increase the entanglement of the system only by local operations and classical communications [14], we have to prepare the needed entangled states in one location, and then distribute the entanglement among several distant locations. During the transmission, storage and processing, the entanglement of the state will unavoidably decrease because of noises. Furthermore the decrease is exponential to the distance. To achieve a faithful transmission of unknown states, we must, first, purify the noisy quantum channel. Bennett *et al* have proposed the first quantum purification scheme, which can purify some near perfectly entangled pairs out of a large supply of mixed entangled pairs using local operations, such as unilateral Pauli rotations, Bilateral rotations and quantum-XOR operations[15]. The basic steps of the entanglement purification are the operations using C-NOT gate or other logic gates. But, in experiment, the implementation of these logic gates is very difficult. We must find some physical processes to replace the theoretical logic gates. J.W. Pan *et al* have found a linear optical device, polarizing beam splitter, to take the role of the C-NOT gate [16]. Thereafter, some other theoretical and experimental schemes for the entanglement purification have been presented [17-24].

As a review of the previous schemes, we can get a general idea of entanglement purification. The general entanglement purification protocol involves three basic steps: (1) the local general measurements on the total system(including the entangled system and the auxiliary system), (2) the classical communication, (3) postselection: the selection of the entangled pairs with higher purity conditioned on the measurement result of subsystems [25]. Along these steps, J. L. Romero *et al*, recently, proposed a physical scheme to purify the mixed entangled states of cavity modes [18]. We find that the purification of non-maximally entangled states of bipartite system has been researched intensively. But there are few schemes for purifying the non-maximally entangled states of three particles in the literature, such as the non-maximally entangled pure W states[+]. In the previous article[12], we have concentrated the two-atom entangled state using cavity QED techniques. Here we will discuss the distillation of $W'$ state.

W state is a special kind of entangled state in the tripartite system. There is a more robust entanglement in it than in the GHZ state when one of the three particles was traced out [26]. Because of the special property, when W state is used in the quantum communication, there will be some novel results [27]. So it is of practical significance to

---

[+] Rigorously, the state should be written as $W'$, because the W state is in the form:
$|W_3\rangle = \frac{1}{\sqrt{3}}(|001\rangle + |010\rangle + |100\rangle)$, while $|W_3'\rangle = a|001\rangle + b|010\rangle + c|100\rangle$

concentrate the $W'$ state.

The rest of the paper is outlined as follows: in section II, we will discuss the general distillation protocols for $W'$ state, then a physical scheme realizing the general protocol will be discussed in section III, and the last part, section IV, is the conclusion.

## II  The general distillation protocols for $W'$ state

The general form of W state for $N$ particles is:

$$|W_N\rangle = \frac{1}{\sqrt{N}}|N-1,1\rangle, \tag{1}$$

where $|N-1,1\rangle$ is the symmetric state involving $N$-1 zeros and 1 ones. Let $N=3$, we will get the W state:

$$|W_3\rangle = \frac{1}{\sqrt{3}}(|001\rangle + |010\rangle + |100\rangle). \tag{2}$$

Here, we suppose that the preparation scheme is not optimal and thus the three particles are initially prepared in the $W'$ state in the form:

$$|W_3'\rangle = a|001\rangle + b|010\rangle + c|100\rangle. \tag{3}$$

Without loss of generality, we suppose that the coefficients $a$, $b$, c are real numbers, $a^2 + b^2 + c^2 = 1$, $a \geq b \geq c$, and all known for us. Where the subscripts denote the particles 1, 2 and 3. Assume that the three particles are shared by two distant users A and B. A has the access to particle 1 and B has access to particles 2, 3. Then the non-maximally entangled pure state can be concentrated by local operations. After concentrated, one particle of Bob's particles 2, 3 can be sent to the third user Cliff to construct a three-user quantum channel with high entanglement.

To extract W state from $W'$ state, we will introduce an auxiliary particle, which is initially prepared in the state $|0\rangle_a$. Under the basis $\{|0\rangle_3|0\rangle_a, |1\rangle_3|0\rangle_a, |0\rangle_3|1\rangle_a, |1\rangle_3|1\rangle_a\}$, we will operate a joint unitary transformation on the particle 3 and the auxiliary particle. The unitary transformation, which is in the form:

$$U_1 = \begin{pmatrix} 1 & 0 & 0 & 0 \\ 0 & \dfrac{c}{a} & \sqrt{1-\dfrac{c^2}{a^2}} & 0 \\ 0 & -\sqrt{1-\dfrac{c^2}{a^2}} & \dfrac{c}{a} & 0 \\ 0 & 0 & 0 & 1 \end{pmatrix}, \quad (4)$$

will lead to the following evolution:

$$(a|001\rangle_{123} + b|010\rangle_{123} + c|100\rangle_{123})|0\rangle_a$$

$$\xrightarrow{U_1} (c|001\rangle_{123} + b|010\rangle_{123} + c|100\rangle_{123})|0\rangle_a - \sqrt{a^2-c^2}\,|000\rangle_{123}|1\rangle_a. \quad (5)$$

From the state in (5), we get that the $W'$ state has not been completely concentrated, provided the fact that the auxiliary particle is in the $|0\rangle_a$ state. So, at this moment, we will not measure the auxiliary particle. Instead, we will perform another joint unitary transformation on particle 2 and the auxiliary particle. This time, the transformation takes a new expression:

$$U_2 = \begin{pmatrix} 1 & 0 & 0 & 0 \\ 0 & m & \sqrt{1-m^2} & 0 \\ 0 & -\sqrt{1-m^2} & m & 0 \\ 0 & 0 & 0 & 1 \end{pmatrix}, \quad (6)$$

where the basis under which the transformation is constructed is $\{|0\rangle_2|0\rangle_a, |1\rangle_2|0\rangle_a, |0\rangle_2|1\rangle_a, |1\rangle_2|1\rangle_a\}$, and $m = \dfrac{bc \pm \sqrt{(a^2-c^2)(1-3c^2)}}{1-2c^2}$. Through the second transformation (6), the state of the total system expressed in (5) will undergo the following evolution:

$$(c|001\rangle_{123} + b|010\rangle_{123} + c|100\rangle_{123})|0\rangle_a - \sqrt{a^2-c^2}\,|000\rangle_{123}|1\rangle_a$$

$$\xrightarrow{U_2} c(|001\rangle_{123} + |010\rangle_{123} + |100\rangle_{123})|0\rangle_a - \left(b\sqrt{1-m^2} + m\sqrt{a^2-c^2}\right)|000\rangle_{123}|1\rangle_a. \quad (7)$$

Then a measurement will be operated on the auxiliary particle. If the auxiliary particle is in the $|0\rangle_a$ state, we have extracted the W state from $W'$ state:

$$c \times \sqrt{3} \times \frac{1}{\sqrt{3}}(|001\rangle_{123} + |010\rangle_{123} + |100\rangle_{123}), \tag{8}$$

and the success probability is :

$$P = 3c^2. \tag{9}$$

If the auxiliary particle is in the $|1\rangle_a$ state, we could not extract a W state from the $W'$ state, namely, the distillation fails. Next we will give another distillation protocol, which is more robust than the first one.

In this protocol, similarly, we will give a unitary transformation on particle 3 and the auxiliary particle initially prepared in $|0\rangle_a$ state:

$$U'_1 = \begin{pmatrix} 1 & 0 & 0 & 0 \\ 0 & \frac{c}{a} & 0 & \sqrt{1-\frac{c^2}{a^2}} \\ 0 & 0 & -1 & 0 \\ 0 & \sqrt{1-\frac{c^2}{a^2}} & 0 & -\frac{c}{a} \end{pmatrix}, \tag{10}$$

which will lead to the transformation:

$$(a|001\rangle_{123} + b|010\rangle_{123} + c|100\rangle_{123})|0\rangle_a$$

$$\xrightarrow{U'_1} (c|001\rangle_{123} + b|010\rangle_{123} + c|100\rangle_{123})|0\rangle_a + \sqrt{a^2-c^2}\,|001\rangle_{123}|1\rangle_a, \tag{11}$$

Subsequently, we will perform another unitary transformation on particle 2 and the auxiliary particle:

$$U'_2 = \begin{pmatrix} 1 & 0 & 0 & 0 \\ 0 & \frac{c}{b} & 0 & \sqrt{1-\frac{c^2}{b^2}} \\ 0 & 0 & -1 & 0 \\ 0 & \sqrt{1-\frac{c^2}{b^2}} & 0 & -\frac{c}{b} \end{pmatrix}, \tag{12}$$

Then the evolution of the total system can be expressed as:

$$(c|001\rangle_{123} + b|010\rangle_{123} + c|100\rangle_{123})|0\rangle_a + \sqrt{a^2-c^2}\,|001\rangle_{123}|1\rangle_a$$

$$\xrightarrow{U_2'} c(|001\rangle_{123} + |010\rangle_{123} + |100\rangle_{123})|0\rangle_a + (\sqrt{b^2-c^2}|010\rangle_{123} - \sqrt{a^2-c^2}|001\rangle_{123})|1\rangle_a. \quad (13)$$

Then we will measure the auxiliary particle. If the result is $|0\rangle_a$, we extract the W state successfully from the $W'$ state with probability $P' = 3c^2$. If the result is $|1\rangle_a$, at the first sight, the distillation fails. But when you have analyzed the collapsed state, you will find that the collapsed state is a product state of a single state of particle 1 and a entangled state of particles 2, 3:

$$\sqrt{b^2-c^2}|010\rangle_{123} - \sqrt{a^2-c^2}|001\rangle_{123} = (\sqrt{b^2-c^2}|10\rangle_{23} - \sqrt{a^2-c^2}|01\rangle_{23})|0\rangle_1, \quad (14)$$

If $a=b$, the state of the particles 2,3 will collapse into the maximally entangled state directly:

$$\sqrt{2(b^2-c^2)} \times \frac{1}{\sqrt{2}}(|10\rangle_{23} - |01\rangle_{23}), \quad (15)$$

The success probability is $P_1' = 2(b^2-c^2)$. At this moment, we get a maximally entangled state without any further operations. If $a \neq b$, we still can extract a two-particle maximally entangled state from it by performing another unitary transformation [13] on particle 3 and another auxiliary particle initially prepared in the $|0\rangle_a$ state. We give the general form of the transformation:

$$U_3' = \begin{pmatrix} 1 & 0 & 0 & 0 \\ 0 & \frac{\sqrt{b^2-c^2}}{\sqrt{a^2-c^2}} & 0 & \sqrt{1-\left(\frac{\sqrt{b^2-c^2}}{\sqrt{a^2-c^2}}\right)^2} \\ 0 & 0 & -1 & 0 \\ 0 & \sqrt{1-\left(\frac{\sqrt{b^2-c^2}}{\sqrt{a^2-c^2}}\right)^2} & 0 & -\frac{\sqrt{b^2-c^2}}{\sqrt{a^2-c^2}} \end{pmatrix}. \quad (16)$$

Resulting from the transformation, the state of the particles 2, 3 and the auxiliary particle will undergo the evolution:

$$(\sqrt{b^2-c^2}|10\rangle_{23} - \sqrt{a^2-c^2}|01\rangle_{23})|0\rangle_a$$

$$\xrightarrow{U_3'} \sqrt{b^2-c^2}(|10\rangle_{23} - |01\rangle_{23})|0\rangle_a - \sqrt{a^2-b^2}|01\rangle_{23}|1\rangle_a. \quad (17)$$

If the auxiliary particle is measured in the state $|0\rangle_a$, the non-maximally entangled pure state of particles 2,3 is concentrated with probability $P_2' = 2(b^2 - c^2)$; If the auxiliary particle is in the $|1\rangle_a$ state, the distillation fails. The distillation for two-particle entangled states can be realized using cavity QED techniques [12]. Then Bob can send one of the particles to Cliff, so a quantum channel has been constructed between Bob and Cliff despite the failure of distillation for W state.

During the second distillation protocol, we get two different results. At first sight, after the second unitary transformation, the state of three particles can collapse into a W state conditioned on the fact that the result of measurement on auxiliary particle is $|0\rangle_a$. Then if we get the result $|1\rangle_a$, the distillation fails, that is to say, the state in (14) is "garbage". Although the collapsed state is unentangled for particles 1, 2, 3, particles 2, 3 are still entangled provided the 1 particle is traced out. Through another transformation we can get a concentrated bipartite entangled state.

After distillation for W state, one particle of Bob's two particles can be sent to Cliff to construct a quantum channel with high entanglement among the three users. Here, we have made use of the "garbage", and minimize the entanglement waste inherent in the distillation process.

After giving the general entanglement distillation protocols, we will present a feasible physical scheme, which can realize the first general distillation protocol via cavity QED techniques.

## III  The physical scheme for the distillation of $W'$ state

Suppose the non-maximally entangled state of the three atoms is in the form:

$$|W_3'\rangle_{123} = a|gge\rangle_{123} + b|geg\rangle_{123} + c|egg\rangle_{123}, \tag{18}$$

where $a^2 + b^2 + c^2 = 1$, the subscripts denote the two-level atoms 1, 2 and 3. Without loss of generality, we can assume $a \geq b \geq c$. $|e\rangle$, $|g\rangle$ are the excited and ground state of the atoms respectively. The preparation of this kind of states can be realized by non-linear interaction between atoms and other systems [28-30].

Assume that the three atoms have been distributed among two distant locations, denoted by Alice and Bob respectively. Bob has the access to atoms 2, 3, and Alice has the access to atom 1. After concentration, one of Bob's two atoms can be sent to Cliff to construct a quantum channel with high entanglement among the three users.

To concentrate the non-maximally entangled pure state $W'$, we must introduce an ancillary system, which is a high fineness cavity at Bob's location. Bob should prepare the cavity in the vacuum state $|0\rangle_c$ (the subscript $c$ denotes the high fineness cavity) initially, and a detector of single photon must, at the same time, be in the access of Bob. The total system is depicted in Fig.1.

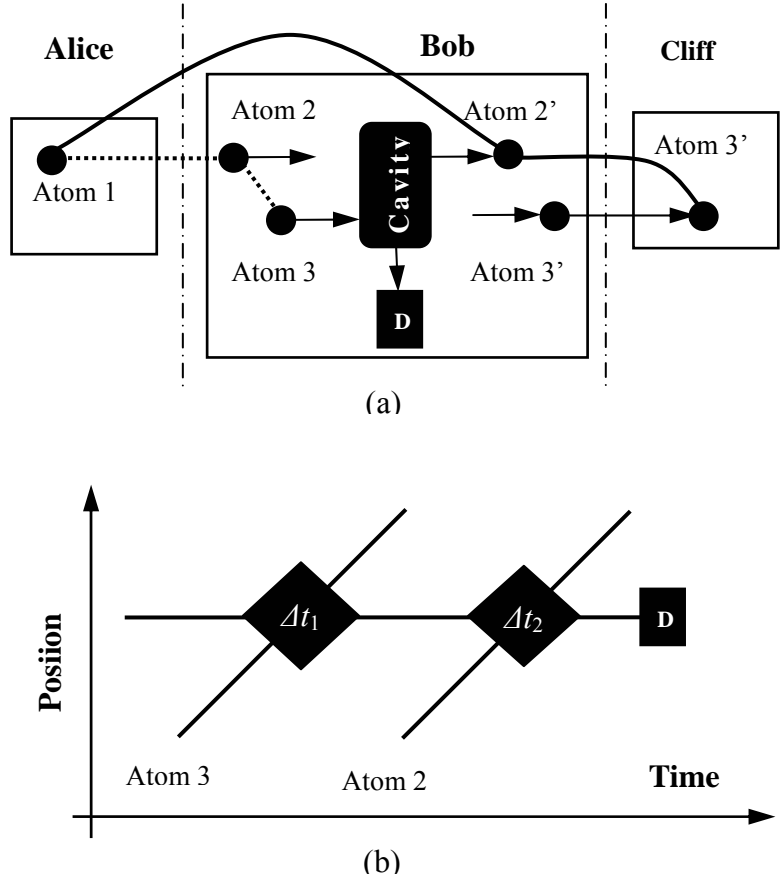

(a)

(b)

Fig.1. Schematic diagram of the entanglement distillation for $W'$ state. (a) the main configuration of the scheme. The broken line denotes the entanglement of the $W'$ state, and the bold line denotes the entanglement of W state. (b) in Bob' location, two atoms are sent through the same cavity in turn. D denotes the detector.

Firstly, Bob will send the atom 3 through the cavity. In the cavity, the atom 3 will interact with the cavity field. In the Jaynes-Cummings model for the interaction between a two-level atom and a single mode field, the Hamiltonian of the system can be expressed as:

$$\hat{H} = \omega a^+ a + \omega_0 S_z + \varepsilon(aS_+ + a^+ S_-), \tag{19}$$

where $\omega_0$ is the atomic transition frequency and $\omega$ is the cavity mode frequency, $a, a^+$ denote the annihilation and creation operators of the cavity mode, $S_+, S_-$ and $S_z$ are atomic operators, $S_+ = |e\rangle\langle g|$, $S_- = |g\rangle\langle e|$, $S_z = \frac{1}{2}(|e\rangle\langle e| - |g\rangle\langle g|)$; $\varepsilon$ is the coupling constant between atom and cavity mode. Here we can modulate the frequency of the cavity mode so that the interaction is a resonant one.

After an interaction time $\Delta t_1$, the evolution of the state of the total system is as follows:

$$\left(a|gge\rangle_{123} + b|geg\rangle_{123} + c|egg\rangle_{123}\right)|0\rangle_c$$

$$\xrightarrow{U(\Delta t_1)} e^{-\frac{i}{2}\omega\Delta t_1} a\cos\varepsilon\Delta t_1 |gge\rangle_{123}|0\rangle_c + e^{\frac{i}{2}\omega\Delta t_1}\left(b|geg\rangle_{123} + c|egg\rangle_{123}\right)|0\rangle_c$$

$$- ie^{-\frac{i}{2}\omega\Delta t_1} a\sin\varepsilon\Delta t_1 |ggg\rangle_{123}|1\rangle_c. \tag{20}$$

When the atom 3 has been out of the cavity, Bob will send the atom 2 through the same cavity. Similarly, we suppose the atomic transition frequency is resonant with that of the cavity mode. Assuming that the interaction time is $\Delta t_2$, we will get the following transformation:

$$e^{-\frac{i}{2}\omega\Delta t_1} a\cos\varepsilon\Delta t_1 |gge\rangle_{123}|0\rangle_c + e^{\frac{i}{2}\omega\Delta t_1}\left(b|geg\rangle_{123} + c|egg\rangle_{123}\right)|0\rangle_c$$

$$- i\, e^{-\frac{i}{2}\omega\Delta t_1} a\sin\varepsilon\Delta t_1 |ggg\rangle_{123}|1\rangle_c$$

$$\xrightarrow{U(\Delta t_2)} \left[e^{\frac{i}{2}\omega(\Delta t_2 - \Delta t_1)} a\cos\varepsilon\Delta t_1 |gge\rangle_{123} + e^{\frac{i}{2}\omega(\Delta t_2 + \Delta t_1)} c|egg\rangle_{123}\right.$$

$$\left. + \left(e^{-\frac{i}{2}\omega(\Delta t_2 - \Delta t_1)} b\cos\varepsilon\Delta t_2 - e^{-\frac{i}{2}\omega(\Delta t_2 + \Delta t_1)} a\sin\varepsilon\Delta t_1 \sin\varepsilon\Delta t_2\right)|geg\rangle_{123}\right]|0\rangle_c$$

$$-i\left(e^{-\frac{i}{2}\omega(\Delta t_2-\Delta t_1)}b\sin\varepsilon\Delta t_2+e^{-\frac{i}{2}\omega(\Delta t_2+\Delta t_1)}a\sin\varepsilon\Delta t_1\cos\varepsilon\Delta t_2\right)|ggg\rangle_{123}|1\rangle_c \quad (21)$$

After interaction, Bob will detect the cavity field. If we select the optimal interaction times:

$$\Delta t_1=\frac{1}{\varepsilon}\cos^{-1}\frac{c}{a}, \quad (22)$$

$$\Delta t_2=\frac{1}{\varepsilon}\left(\sin^{-1}\frac{b}{\sqrt{1-2c^2}}-\sin^{-1}\frac{c}{\sqrt{1-2c^2}}\right), \quad (23)$$

we can get the W state provided the fact that the detector's result is $|0\rangle_c$, namely, there is no photon in the cavity. The form of the concentrated state is:

$$c\times\sqrt{3}\times\frac{1}{\sqrt{3}}(|gge\rangle_{123}+|geg\rangle_{123}+|egg\rangle_{123}), \quad (24)$$

where we have discarded the phase factor. So we can get W state with probability $P=3c^2$. That is to say, the success probability is only determined by the smaller coefficient of the superposition state to be concentrated. If the detector detected a photon in the cavity, the distillation fails. When the $W'$ state has been concentrated, one of Bob's two atoms will be sent to the Cliff, that is to say, there exists a high entanglement quantum channel between Alice Bob and Cliff, which becomes a more robust resource in the three-user quantum communication [27].

Although the protocol looks like the two-partite-distillation protocols [31], it also can be looked as the three-partite-distillation protocol if we make some auxiliary operations. Here, if the distance between Bob and Cliff is short enough, one of the Bob's two atoms can be sent to Cliff physically to construct a three-user maximally entangled quantum channel after distillation. On the other hand, if the distance between Bob and Cliff is too long, Bob can make use of one ebit (bipartite maximally entangled state) to teleport the state of one of his two atoms to Cliff after concentration. Then the entanglement of W state has been distributed among three distant users.

## IV    Conclusion

In conclusion, we have presented two different general entanglement distillation protocols for non-maximally entangled pure W states, and the success probabilities are all dependent on the smaller coefficient of the superposition of the $W'$ state. In the process, we introduced an auxiliary system. To concentrate the non-maximally entangled pure $W'$ state, we proposed two kinds of unitary transformations. For the first case, we also gave the corresponding physical scheme based on cavity QED techniques. The second case involves some special features, that is to say, after transformation, when the measurement result on the auxiliary particle tells us that the distillation succeeds, we get the tripartite W state. If the result tells us that the distillation fails, the state of the three particles collapses into the state from which the W state can not be extracted. It is an unentangled state for the three particles, but an entangled state for the particles 2, 3. So we can get a maximally entangled bipartite state from it. So, adopting the transformation, we can make a full use of the entanglement source. But we have not found a physical scheme to realize the transformation for the second case, which is a subject to be further researched. In addition, we haven't discussed the more general case, purification of mixed states.

The current paper is mainly about the probabilistic concentration of single copy of non-maximally entangled pure W state, which is different from the scheme proposed by Bennett et al[23] involving the entanglement concentration from many non-maximally entangled pure states. The common advantage of the two general protocols is that we need to measure the auxiliary system only once, which decreases the error rate resulting from the imperfect quantum operations, and the physical realization of the general protocol becomes a bridge between theoretical protocols and experimental ones.

## Acknowledgements

This work is supported by the Natural Science Foundation of Anhui Province under Grant No: 03042401 and the Natural Science Foundation of the Education Department of Anhui Province under Grant No: 2002kj026, also by the fund of the Core Teacher of Ministry of National Education under Grant No: 200065.